\documentclass[longauth, traditabstract]{aa} 
\usepackage{graphicx}
\usepackage{txfonts}
\usepackage[numbers]{natbib}
\begin{document}
\bibpunct{[}{]}{,}{n}{}{;}

\title{{\it Herschel} observations of extra-ordinary sources: Detection of
  Hydrogen Fluoride in absorption towards Orion~KL\thanks{{\it Herschel} is an ESA space
    observatory with science instruments provided by European-led
    Principal Investigator consortia and with important participation
    from NASA.}}

\author{
T.~G.~Phillips,\inst{1},
E.~A.~Bergin,\inst{2}
D.~C.~Lis,\inst{1}
D.~A.~Neufeld,\inst{3}
T.~A.~Bell,\inst{1}
S.~Wang,\inst{2}
N.~R.~Crockett,\inst{2}
M.~Emprechtinger,\inst{1}
G.A. Blake\inst{1}
E.~Caux,\inst{4,5}
C.~Ceccarelli,\inst{6}
J.~Cernicharo,\inst{7}
C.~Comito,\inst{8}
F.~Daniel,\inst{7,9}
M.-L.~Dubernet,\inst{10,11}
P.~Encrenaz,\inst{9}
M.~Gerin,\inst{9}
T.~F.~Giesen,\inst{12}
J.~R.~Goicoechea,\inst{7}
P.~F.~Goldsmith,\inst{13}
E.~Herbst,\inst{14}
C.~Joblin,\inst{4,5}
D.~Johnstone,\inst{15}
W.~D.  Langer\inst{13}
W.D.~Latter\inst{16} 
S.~D.~Lord,\inst{16}
S.~Maret,\inst{6}
P.~G.~Martin,\inst{17}
G.~J.~Melnick,\inst{18}
K.~M.~Menten,\inst{8}
P.~Morris,\inst{16}
H.~S.~P. M\"uller,\inst{12}
J.~A.~Murphy,\inst{19}
V.~Ossenkopf,\inst{12,20}
J.~C.~Pearson,\inst{13}
M.~P\'erault,\inst{9}
R.~Plume,\inst{21}
S.-L.~Qin,\inst{12}
P.~Schilke,\inst{8,12}
S.~Schlemmer,\inst{12}
J.~Stutzki,\inst{12}
N.~Trappe,\inst{19}
F.~F.~S.~van der Tak,\inst{21}
C.~Vastel,\inst{4,5}
H.~W.~Yorke,\inst{13}
S.~Yu,\inst{13}
J.~Zmuidzinas,\inst{1}
A.~Boogert,\inst{16}
R.~G\"usten,\inst{8}
P.~Hartogh,\inst{22}
N.~Honingh,\inst{12}
A.~Karpov,\inst{1}
J.~Kooi,\inst{1}
J.-M.~Krieg,\inst{9}
\and
R.~Schieder\inst{12}
}
\institute{California Institute of Technology, Cahill Center for Astronomy and Astrophysics 301-17, Pasadena, CA 91125 USA\\
             \email{tgp@submm.caltech.edu}
\and Department of Astronomy, University of Michigan, 500 Church Street, Ann Arbor, MI 48109, USA 
\and  Department of Physics and Astronomy, Johns Hopkins University, 3400 North Charles Street, Baltimore, MD 21218, USA
\and Centre d'\'etude Spatiale des Rayonnements, Universit\'e de Toulouse [UPS], 31062 Toulouse Cedex 9, France
\and CNRS/INSU, UMR 5187, 9 avenue du Colonel Roche, 31028 Toulouse Cedex 4, France
\and Laboratoire d'Astrophysique de l'Observatoire de Grenoble, 
BP 53, 38041 Grenoble, Cedex 9, France.
\and Centro de Astrobiolog\'ia (CSIC/INTA), Laboratiorio de Astrof\'isica Molecular, Ctra. de Torrej\'on a Ajalvir, km 4
28850, Torrej\'on de Ardoz, Madrid, Spain
\and Max-Planck-Institut f\"ur Radioastronomie, Auf dem H\"ugel 69, 53121 Bonn, Germany 
\and LERMA, CNRS UMR8112, Observatoire de Paris and \'Ecole Normale Sup\'erieure, 24 Rue Lhomond, 75231 Paris Cedex 05, France
\and LPMAA, UMR7092, Universit\'e Pierre et Marie Curie,  Paris, France
\and  LUTH, UMR8102, Observatoire de Paris, Meudon, France
\and I. Physikalisches Institut, Universit\"at zu K\"oln,
              Z\"ulpicher Str. 77, 50937 K\"oln, Germany
\and Jet Propulsion Laboratory,  Caltech, Pasadena, CA 91109, USA
\and Departments of Physics, Astronomy and Chemistry, Ohio State University, Columbus, OH 43210, USA
\and National Research Council Canada, Herzberg Institute of Astrophysics, 5071 West Saanich Road, Victoria, BC V9E 2E7, Canada 
\and Infrared Processing and Analysis Center, California Institute of Technology, MS 100-22, Pasadena, CA 91125
\and Canadian Institute for Theoretical Astrophysics, University of Toronto, 60 St George St, Toronto, ON M5S 3H8, Canada
\and Harvard-Smithsonian Center for Astrophysics, 60 Garden Street, Cambridge MA 02138, USA
\and  National University of Ireland Maynooth. Ireland
\and SRON Netherlands Institute for Space Research, PO Box 800, 9700 AV, Groningen, The Netherlands
\and Department of Physics and Astronomy, University of Calgary, 2500
University Drive NW, Calgary, AB T2N 1N4, Canada
\and MPI f\"ur Sonnensystemforschung, D 37191 Katlenburg-Lindau,
Germany
}


\abstract{We report a detection of the fundamental rotational
  transition of hydrogen fluoride in absorption towards Orion~KL using
  {\it Herschel}/HIFI. 
After the removal of contaminating features associated with common molecules ("weeds"),
the HF spectrum shows a P-Cygni profile, with
weak redshifted emission and strong blue-shifted absorption,
associated with the low-velocity molecular outflow.
We derive an estimate of $2.9 \times 10^{13}\,\rm cm^{-2}$ for the HF
column density responsible for the broad absorption component.  
Using our best estimate of the H$_2$ column density within the low-velocity molecular outflow, we obtain a lower limit of $\sim 1.6 \times
10^{-10}$ for the HF abundance relative to hydrogen nuclei, corresponding to $\sim$ 0.6\% of the solar abundance of fluorine. This value is close to that inferred from previous ISO observations of HF J=2--1 absorption towards Sgr B2, but is in sharp contrast to the lower limit of $6 \times 10^{-9}$ derived by Neufeld et al. (2010) for cold, foreground clouds on the line of sight towards G10.6-0.4.
}

   \titlerunning{Hydrogen fluoride in Orion}
	\authorrunning{Phillips et al.}
   \maketitle
%

\section{Introduction}

Hydrogen fluoride (HF) is expected to be the main reservoir of fluorine in
the interstellar medium because  it is easily produced by the
exothermic reaction of F with H$_2$ (Neufeld, 
Wolfire \& Schilke 2005; Neufeld \& Wolfire 2009) and its very strong chemical bond
makes this molecule relatively insensitive to UV photodissociation.
Interstellar HF was first detected by Neufeld et al. (1997) with the
Infrared Space Observatory (ISO).  The $J = 2-1$ rotational transition was observed in
absorption towards Sgr B2, at a low spectral resolution using the ISO long-wavelength spectrometer (LWS). 
The HIFI instrument (de Graauw et al. 2010) aboard the
{\em Herschel Space Observatory} (Pilbratt et al. 2010) has allowed 
observations of the ground state rotational transition of HF at 1.232~THz to be 
performed for the first time, at high spectral resolution.
This transition is expected
to be generally observed in absorption because of the very large A
coefficient (e.g. Neufeld et al. 2010). 
Only extremely dense regions could possibly generate enough collisional 
excitation to yield an HF feature with a positive frequency-integrated flux.

A full HIFI spectral scan of band 5a, with frequency coverage from
1.109 to 1.239~THz, was carried out as part of the Guaranteed Time Key
Program \emph{Herschel Observations of Extraordinary Sources: The
  Orion and Sagittarius B2 Star-forming Regions} (HEXOS). With a strong
continuum, it
might be expected that Orion would exhibit numerous absorption lines;
however it is well known to exhibit no absorption lines at mm
wavelengths.   For instance even CO $J= 1- 0$, which is seen
with self-reversals towards many star-forming regions, has a smooth
emission line profile with no self-absorption (e.g. Tauber et al.
1991). The lack of absorption has been attributed to competing
excitation gradients (external heating from $\theta^1C$ and internal
heating from the embedded massive protostars), the location of the
H\,{\sc ii} region
on the front of the cloud, and the presence of
numerous unresolved dense ($n_{H_2} > 10^5$ cm$^{-3}$) clumps along
the line of sight (Tauber et al. 1991).   At shorter wavelengths some
evidence for absorption is found.  Betz \& Boreiko (1989) find that the
blue side of the 
fundamental rotational transition of OH is completely absorbed, with
only a red-shifted emission component.    The far-infrared survey of Lerate et al. (2006)
with ISO-LWS (188 to 44~$\mu$m), has shown that the shapes of
water and OH lines gradually change from pure emission at the longest
wavelengths to mostly P-Cygni profiles at the shortest wavelengths.  At
even shorter mid-infrared wavelengths, strong water absorption has been
detected by Wright et al. (2000) using ISO-SWS. With the $\sim 10$
km\,s$^{-1}$ resolution of the SWS Fabry Perot, the absorption is shown
to be blue-shifted with respect to the source systemic velocity,
extending down to about $-50$ km\,s$^{-1}$. In recent observations
with the CRIRES spectrograph on the Very Large Telescope, Beuther et
al.\ (2010) have obtained spectra of the R-branch of the $\varv = 1-0$ band
of both $^{13}$CO and $^{12}$CO toward the BN object and ``source n''
within Orion~KL.  These spectra, which are consistent with earlier
spectra obtained at lower resolution and signal-to-noise ratio by
Scoville et al.\ (1983), also show strong absorption at velocities
between about $-25$ and 12 km\,s$^{-1}$. As we will discuss below, the
particular excitation of these lines (ground rotational state HF and
ground vibrational state CO) makes them strong 
candidates to be seen in absorption, provided favorable geometry and
strong background continuum exist. Ground state rotational lines of
water isotopologues have similar excitation requirements and HIFI
observations of $para-$H$_2^{18}$O and $para-$H$_2^{17}$O, observed
separately in band 4b, are also discussed here.

\section{Observations}

HIFI observations presented here were carried out on March 6, 2010
using the dual beam switch (DBS) mode pointed towards the Orion Hot
Core at $\alpha_{J2000} = 5^h35^m14.3^s$ and $\delta_{J2000} =
-5^{\circ}22'36.7''$. The DBS reference beams lie approximately
3$^{\prime}$ east and west (i.e. perpendicular to the orientation of
the Orion Molecular Ridge; e.g. Ungerechts et al. 1997). We used the
Wide Band Spectrometer providing a spectral resolution of 1.1 MHz
(0.26 km s$^{-1}$) over a 4~GHz IF bandwidth. Although both H and V
polarization data were obtained, we only present here data from the H
polarization, reduced using HIPE (Ott 2010) with pipeline version 2.4.

The band 4b and 5a spectral scans consist of double sideband
spectra with a redundancy of 6, which gives observations of a
lower or upper sideband frequency with 6 different frequency settings of
the local oscillator. This allows for the deconvolution and isolation
of a single sideband spectrum (Comito \& Schilke 2002). We applied the standard
HIFI deconvolution using the {\it doDeconvolution} task within HIPE.
All data presented here are deconvolved single sideband
spectra, including the continuum.
Regions of the spectrum free of
lines were isolated and give a typical rms of $T_A^* =$ 0.17~K at the
original spectral resolution. The HIFI beam size at 1.23~THz is
17$^{\prime\prime}$, with an assumed main beam efficiency of 0.67.

A $^{13}$CO $J = 2-1$ spectrum towards Orion~BN/KL was also obtained
using the facility receiver and spectrometers of the Caltech
Submillimeter Observatory (CSO) atop Mauna Kea in Hawaii. The CSO beam
size at 220 GHz is $\sim 33^{\prime\prime}$ and the main beam
efficiency is $\sim 0.68$.

   %
   
%
 %

\section{Results}

\subsection{First detection of submm absorption towards Orion}

\begin{figure}
   \centering
   \includegraphics[width=0.95\columnwidth]{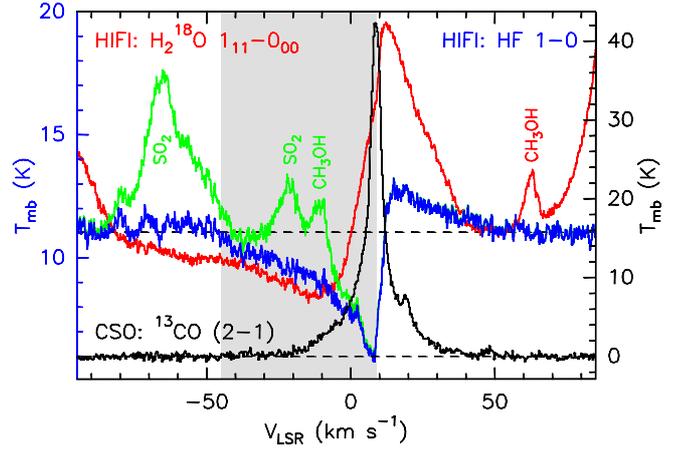}
   \caption{HIFI detection of interstellar 
      HF $J = 1-0$ (rest frequency 1232.47627 GHz; Nolt et al. 1987) and 
     $para-$H$_2^{18}$O 1$_{11} - 0_{00}$ (at 1101.698256 GHz) towards
     Orion~BN/KL (blue and red, respectively). The region of
     low-velocity HF absorption is highlighted in grey. Both
     observations are from HIFI band 5a and are referenced to the
     temperature scale on the left. 
     CSO $^{13}$CO  $J = 2-1$ spectrum towards the same
     position is shown in black, using the temperature scale on the right. The HF
     absorption is blended with the emission of CH$_3$OH 
     ($\sim -10$ km\,s$^{-1}$) and SO$_2$  
($\sim -20$ and $-65$~km\,s$^{-1}$), shown in green. Subtraction of these
     contaminating lines results in the dark blue
     HF spectrum. }
         \label{fig1}
\end{figure}

Fig.~\ref{fig1} shows the detection of HF $J = 1-0$ and $para-$H$_2^{18}$O
1$_{11} - 0_{00}$ in emission and absorption towards Orion BN/KL (blue
and red histograms, respectively). Both lines show 
high-velocity emission line wings on the red side of the systemic velocity of 9
km\,s$^{-1}$, a sharp drop near the systemic velocity, and broad
absorption extending towards lower (blue) velocities. In
contrast, $^{13}$CO $J = 2-1$ emission (black histogram) shows broad line wings
superposed on a narrow feature at 9 km\,s$^{-1}$, but no evidence for
absorption. 
  
There are a number of issues which must be addressed. First these data were
obtained using DBS which places the reference beams 3$^{\prime}$ away
from the central hot core. This is large enough to avoid any
reference position contamination from the hot core and shock, but
might encompass emission from the extended molecular ridge. Both HF
and $para-$H$_2^{18}$O have high dipole moments and fast 
($>10^{-2}$
s$^{-1}$) spontaneous de-excitation rates leading to critical densities
in excess of 10$^{8}$ cm$^{-3}$ (Reese et al.\ 2005; 
Grosjean, Dubernet \& Ceccarelli 2003).   Beyond the hot core, the density is
well below this value (Bergin et al. 1996) and, for HF, extended emission is
unlikely.    This may not be the case for $para-$H$_2^{18}$O, as the ground state emission of $ortho$-H$_2$O is
strong and extended (Snell et al. 2000; Olofsson et al.
2003). Because the DBS mode alternates between two reference
positions, we have used the Level 1 data to compute a difference
spectrum between the two reference positions; we see no evidence for
emission or absorption in such a difference spectrum for either line.   It is very unlikely that
the same level of emission or absorption would be present in the two
reference beams, separated by 6$^{\prime}$ on the sky. In addition,
the extended emission component in Orion is centered at 9 km\,s$^{-1}$
and has a narrow line width of 2--5 km\,s$^{-1}$. For HF we do see 
absorption at the systemic velocity, but also a broad blue-shifted
absorption. We thus conclude that the observed absorption is real, and
not an artifact of the observing mode employed.

\subsection{Contamination by interfering lines}

\begin{figure}
   \centering
   \includegraphics[width=1.02\columnwidth]{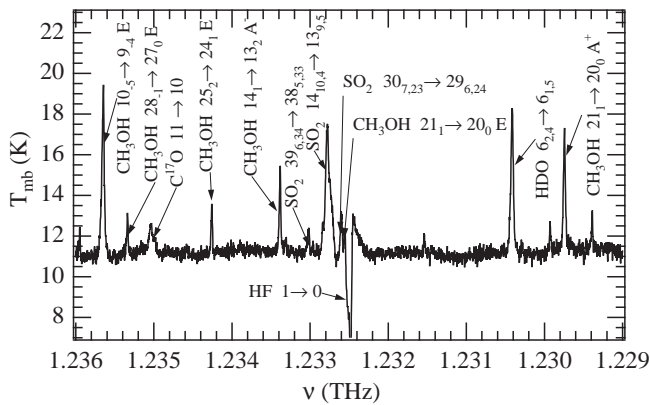}
   \caption{A more extensive look at the molecular spectrum
     surrounding HF $J=1-0$. Transitions of CH$_3$OH and SO$_2$ were
     isolated using the frequencies taken from M\"uller \& Br\"unken (2005)
     and Xu et al. (2008).   Three lines are 
     blended with (or provide background
     to) the HF absorption. 
}
         \label{fig2}
\end{figure}

In the case of HF there is an additional complication in that the
low-velocity blue absorption is blended with emission from CH$_3$OH
$J= 21_1 - 20_0$ E and SO$_2$ $J= 30_{7,23} - 29_{6,24}$ (see grey
area in Fig.~\ref{fig1}). A more extensive look at the spectrum near
HF $J=1-0$ is provided in Fig.~\ref{fig2}. In this 7~GHz wide region
of the spectrum we see multiple prominent lines of both CH$_3$OH and
SO$_2$. For CH$_3$OH, one of the detected transitions is $J= 21_1 -
20_0$ A$^+$, which has a similar ground state energy and 120\%
stronger line strength than the $J= 21_1 - 20_0$ E transition. Thus we
confirm the CH$_3$OH contamination. 
We have
modeled the extensive emission from SO$_2$ (and CH$_3$OH) seen in Band 5a
assuming LTE (with a correction for optical depth). This analysis also
confirms that SO$_2$ $J= 30_{7,23} - 29_{6,24}$ will emit at
appreciable levels and interfere with the HF absorption.

\begin{figure}
   \centering
   \includegraphics[width=0.95\columnwidth]{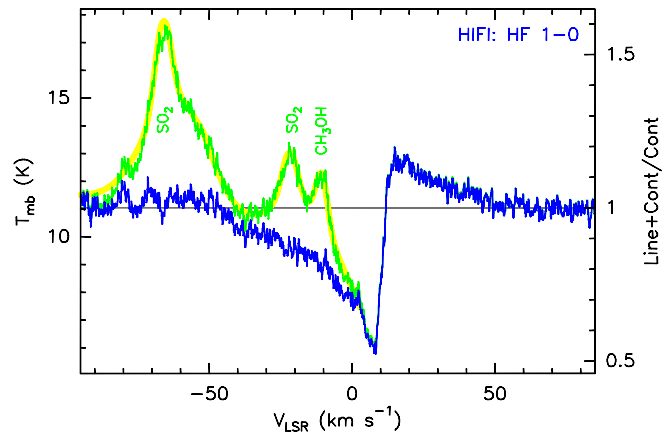}
   \caption{HIFI HF $J=1-0$ spectrum (blue-green histogram) with
     fit to CH$_3$OH and SO$_2$ emission shown in yellow. Subtraction of these
     contaminating lines provides a more complete look at the HF
     absorption (blue histogram). }
         \label{fig3}
\end{figure}

Line contamination by abundant ``weed'' molecules, such as methanol or
SO$_2$, will be a common problem affecting HIFI observations of
molecular hot core sources and modeling tools are being developed to
deal with this problem. The two SO$_2$ lines in the spectrum have been
removed using such a model.\footnote{We made use of the myXCLASS program
  (http://www.astro.uni-koeln.de/projects/schilke/XCLASS), which
  accesses the CDMS (M\"uller at al. 2001, M\"uller et al. 2005; http://www.cdms.de) and JPL
  (Pickett et al. 1998; http://spec.jpl.nasa.gov) molecular data bases.}
In the case of CH$_3$OH, the LTE model is not accurate enough to deal with
the problem and we removed the interfering line by fitting a
single-component gaussian and subtracting the emission. The emission
fit and the final HF absorption spectrum are shown in
Fig.~\ref{fig3}. This spectrum is then used 
for the analysis in \S~\ref{sec:disc}. The
HF absorption full-width at zero intensity (FWZI) is $\sim 50$ km\,s$^{-1}$, which is less than
that of $para-$H$_2^{18}$O (FWZI $\sim$80 km\,s$^{-1}$).
For completeness,
in the computation of the line-to-continuum ratio, we also explored
the possibility that the CH$_3$OH and SO$_2$ lines contribute to the
background emission for the absorbing HF gas. This will be reflected
in our error analysis.

The analysis is simpler for H$_2^{18}$O, which shows no
evidence for any contamination within the absorption velocity range.
For this line we have assumed a continuum value based on the level measured
at frequencies adjacent to the water line.


\section{Discussion}
\label{sec:disc}

After the removal of features attributed to CH$_3$OH and SO$_2$, the
HF $J=1-0$ spectrum shows 
a P-Cygni profile with a broad, blueshifted absorption at LSR
velocities between about $-45$ and 9~km\,s$^{-1}$ and a redshifted emission
component at LSR velocities in the range 12 to 50 km\,s$^{-1}$.
The analysis of the HF emission,
along with other spectral lines detected in Orion, will be
discussed in a future paper. The HF spectrum is strikingly similar to
that of another transition with an extremely high critical density:
the CO fundamental vibrational band (Beuther et al. 2010) which shows
the same combination of absorption at LSR velocities between about
$-25$ and 12~km$\,\rm s^{-1}$ and weak emission extending to $V_{\rm
  LSR} \sim 30$~km$\,\rm s^{-1}$. We suggest that exactly the same
physical processes give rise to the CO $\varv = 1-0$ and HF $J=1-0$ spectra:
outflowing material surrounding the hot core in Orion absorbs and
re-emits continuum radiation from the central source. The
absorption---occurring in front of the source---naturally gives rise
to a blueshifted absorption feature, while re-emission from the back
side of the outflow provides the weak redshifted emission feature. For
transitions with a very high critical density, the absorption of a
``resonance line photon'' (such as HF $J=1-0$) will inevitably be
followed by re-emission.  Were the continuum source surrounded by a
spherical shell that is small compared to the telescope beam, the
areas of the emission and absorption features would be equal. The
simplest interpretation of the observed HF $J=1-0$ spectrum---which
shows an absorption feature that is stronger than the emission
feature---is that the outflowing material is not entirely encompassed
by the beam so that part of the emission flux is unobserved; this
explanation could be tested by means of mapping observations.

 Most of the material appears to have an outflow velocity $\le 20\,\rm
km\,s^{-1}$ and is likely associated with the ``Low Velocity Flow''
(e.g.\ Genzel \& Stutzki 1989). The $para-$$\rm H_2^{18}O$ and $para-$$\rm
H_2^{17}O$ $1_{11}-0_{00}$ (not shown) lines 
also show absorption at negative LSR
velocities, although the absorption extends further, to LSR velocities
as negative as $\sim -80\,\rm km\,s^{-1}$. This behavior may reflect
the presence of enhanced water abundances in the High Velocity Flow
(Franklin et al. 2008). The $para-$$\rm H_2^{18}O$ (and $para-$$\rm
H_2^{17}O$ $1_{11}-0_{00}$) line profiles are also different from HF in
exhibiting an emission feature that is stronger than the absorption
feature. This behavior must imply that collisional excitation provides
an additional excitation mechanism, and may suggest that the rate
coefficients for excitation of the $para-$water transition are
significantly larger than those for excitation of the HF transition.
To date, the collisional excitation of HF has been computed only in
the case where He is the collision partner (Reese et al.\ 2005); the
rate coefficients thereby derived are indeed an order of magnitude
smaller than those computed for the excitation of $para-$water by H$_2$
(Grosjean, Dubernet \& Ceccarelli 2003), but the rate coefficients for
excitation of HF by H$_2$ might be expected to be larger than those
for excitation by He.

We have determined the column density of absorbing molecules
responsible for the broad blueshifted absorption features. If the
absorbing material is assumed to cover the continuum source, we estimate the
velocity-integrated optical depth HF $J=1-0$ as $11.8 \rm \,
km\,s^{-1}$, which implies an HF column density of $2.9 \times 10^{13}
\rm cm^{-2}$ if all molecules are in the ground-state. Uncertainties
introduced by the need to correct for the SO$_2$ and CH$_3$OH emission
features result in possible errors that we estimate as about $\pm
25\%$. Derived under the same set
of assumptions, the column densities of absorbing $para-$$\rm H_2^{18}O$
and $para-$$\rm H_2^{17}O$ are $1.3 \times 10^{13}$~cm$^{-2}$ and $7
\times 10^{12}$~cm$^{-2}$, respectively. Strictly, these values are
all lower limits, because the source could be partially-covered by
clumps of arbitrarily large optical depth and the strong continuum could lead to some excitation. In addition, the water lines have strong emission which might lie behind the absorbing material.  However, the fact that all
three spectral lines show absorption profiles of similar shapes but
different depths, suggests that the optical depths are not extremely
large.

In order to estimate the molecular abundances implied by these column
densities, we require an estimate of the total column density of
H$_2$. Beuther et al.\ (2010) used the observed strength of the $^{13}$CO
$\varv = 1-0$ absorption features along the sight-line to the BN object,
together with standard assumptions about the $\rm ^{13}CO/^{12}CO$ and
$\rm CO/H_2$ abundance ratios, to derive an H$_2$ column density of
$4.9 \times 10^{22}\, \rm cm^{-2}$ for the outflowing absorbing gas.
An alternative estimate has been obtained by Persson et al.\ (2007)
from observations of {\it emission} from $\rm C^{17}O$ pure rotational
lines; this yields a value $3 \times 10^{23}\, \rm cm^{-2}$ for the
total H$_2$ column density within the Low Velocity Flow, half of which
($1.5 \times 10^{23}$~cm$^{-2}$) would be associated with the blue
outflow lobe. We have estimated the H$_2$ column density independently
from the strength of the broad $^{13}$CO 2--1
emission component. Assuming LVG, 100~K kinetic temperature, an H$_2$
density of $1 \times 10^5$~cm$^{-3}$ (Persson et al. 2007), and a
$^{13}$CO fractional abundance of $2 \times 10^{-6}$ (Dickman 1978),
we derive  N(H$_2$) $= 9 \times 10^{22}$ cm$^{-2}$ in the blue
outflow lobe. This value, intermediate between the Persson et al. and
Beuther et al. estimates, is used in the subsequent calculations (a
factor of 2 estimated uncertainty).

With these assumptions, our observations of the blueshifted absorption
feature imply an HF abundance of $1.6 \times 10^{-10}$ relative to
hydrogen nuclei, corresponding to $\sim 0.6\%$ of the solar abundance
of fluorine (Asplund, Grevesse \& Sauval 2009). This value is in sharp
contrast with the lower limit $N({\rm HF})/N_{\rm H} \ge 6 \times
10^{-9}$ obtained by Neufeld et al.\ (2010) for foreground material
along the sight-line to G10.6-0.4, but close to that inferred ($N({\rm
  HF})/N_{\rm H} = 1.5 \times 10^{-10}$) from previous observations of
HF $J=2-1$ absorption toward Sgr B2 (Neufeld et al.\ 1997). Since the
density in the Low Velocity Flow, $n(\rm {H}_2) \sim 10^5 \, cm^{-3}$
(Persson et al.\ 2007) is two orders of magnitude larger than
the typical density of the foreground material probed in observations
of G10.6-0.4, and since HF is expected to be the dominant reservoir of
fluorine in the gas-phase, this abundance difference reflects a
density-dependent depletion of fluorine nuclei onto grain mantles. A
depletion factor $\sim 10^2$ does not seem unreasonable, in light of
the polar nature of the HF molecule; indeed, the freezing and boiling
points of HF are relatively large (190~K and 293~K respectively at 1
atm.\ pressure), so---as in the case of interstellar water---the
freeze-out of HF can be expected to be quite efficient. 
On the other hand, shocks of velocity greater than $\sim 15 \, \rm
km\,s^{-1}$ are predicted to release grain mantles into the gas phase
as the result of sputtering (Draine 1995).  If high depletions are
indeed attained in gas traveling at outflow velocities up to $\sim 35
\rm km\, s^{-1}$, then our observations suggest that the acceleration
to a given terminal velocity can occur without the material suffering
shocks of equal velocity. 
     
\section{Conclusions}

Our observations of hydrogen fluoride toward Orion-KL have revealed an
unusual absorption feature in the spectrum of this source. 
To our knowledge, this is the first report of a submillimeter spectral line with a net negative flux in this archetypical emission line source.
The unusual behavior of the HF $J=1-0$
transition is a consequence of its extremely large critical density,
and is mirrored by mid-infrared observations of the CO $\varv = 1-0$ band.
Thanks to the high spectral resolution achievable with HIFI, the HF
$J=1-0$ line promises to provide a unique probe of the kinematics
of---and depletion within---absorbing material along the sight-line to
bright continuum sources, and one that is uncomplicated by the
collisionally-excited line emission that is usually present in the
spectra of other transitions. Redshifted HF $J=1-0$ absorption may
also prove to be an excellent tracer of the interstellar medium in the
high-redshift Universe; 
the range of redshifts accessible from ground-based submillimeter 
telescopes is indicated by Neufeld et al.\ 2005 (see their Figure 11).

\begin{acknowledgements}
  HIFI has been designed and built by a consortium of institutes and university departments from across 
Europe, Canada and the United States under the leadership of SRON Netherlands Institute for Space
Research, Groningen, The Netherlands and with major contributions from Germany, France and the US. 
Consortium members are: Canada: CSA, U.Waterloo; France: CESR, LAB, LERMA,  IRAM; Germany: 
KOSMA, MPIfR, MPS; Ireland, NUI Maynooth; Italy: ASI, IFSI-INAF, Osservatorio Astrofisico di Arcetri- 
INAF; Netherlands: SRON, TUD; Poland: CAMK, CBK; Spain: Observatorio Astron—mico Nacional (IGN), 
Centro de Astrobiolog'a (CSIC-INTA). Sweden:  Chalmers University of Technology - MC2, RSS \& GARD; 
Onsala Space Observatory; Swedish National Space Board, Stockholm University - Stockholm Observatory; 
Switzerland: ETH Zurich, FHNW; USA: Caltech, JPL, NHSC.
Support for this work was provided by NASA through an award issued by JPL/Caltech.
CSO is supported by the NSF, award AST-0540882.
\end{acknowledgements}


\begin{thebibliography}{}

\bibitem[Asplund et 
al.(2009)]{2009ARA&A..47..481A} Asplund, M., Grevesse, N., Sauval, A.~J., \& Scott, P.\ 2009, \araa, 47, 481 

\bibitem[Bergin et al.(1996)]{1996ApJ...460..343B} Bergin, E.~A., Snell, 
R.~L., \& Goldsmith, P.~F.\ 1996, \apj, 460, 343 

\bibitem[Betz 
\& Boreiko(1989)]{1989ApJ...346L.101B} Betz, A.~L., \& Boreiko, R.~T.\ 1989, \apjl, 346, L101 



\bibitem[Beuther et al.(2010)]{2010arXiv1001.0650B} Beuther, H., Linz, H., 
Bik., A., Goto, M., \& Henning, T.\ 2010, arXiv:1001.0650 

\bibitem[Comito 
\& Schilke(2002)]{2002A&A...395..357C} Comito, C., \& Schilke, P.\ 2002, \aap, 395, 357 


\bibitem[de Graauw et al. (2010)]{DG} de Graauw, Th.\ et al.\, this volume

\bibitem[Dickman (1987)]{Dickman1978} Dickman, R.L. 1978, ApJS, 37, 407

\bibitem[Draine(1995)]{1995Ap&SS.233..111D} Draine, B.~T.\ 1995, \apss, 233, 111

\bibitem[Franklin et al.(2008)]{2008ApJ...674.1015F} Franklin, J., Snell, 
R.~L., Kaufman, M.~J., Melnick, G.~J., Neufeld, D.~A., Hollenbach, D.~J., 
\& Bergin, E.~A.\ 2008, \apj, 674, 1015 


\bibitem[Genzel 
\& Stutzki(1989)]{1989ARA&A..27...41G} Genzel, R., \& Stutzki, J.\ 1989, \araa, 27, 41 


\bibitem[Grosjean et 
al.(2003)]{2003A&A...408.1197G} Grosjean, A., Dubernet, M.-L., \& Ceccarelli, C.\ 2003, \aap, 408, 1197

\bibitem[Lerate et al. (2006)]{lerate2006} Lerate, M.R., Barlow, M.J.,
  Swinyard, B.M., et al. 2006, MNRAS, 370, 597

\bibitem[Muller \& Brunken 2005]{mb05}  M\"uller, H.S.P., \& Br\"unken, S., 2005, J. Mol. Spectrosc. 232,  213.

\bibitem[M{\"u}ller et al.(2001)]{CDMS_1}
M{\"u}ller, H.~S.~P., Thorwirth, S., Roth, D.~A.,
\& Winnewisser, G.
2001, A\&A, 370, L49

\bibitem[M{\"u}ller et al.(2005)]{CDMS_2}
M{\"u}ller, H.~S.~P., Schl{\"o}der, F., Stutzki, J.,
\& Winnewisser, G.
2005, J. Mol. Struct, 742, 215


\bibitem[Neufeld 
\& Wolfire(2009)]{2009ApJ...706.1594N} Neufeld, D.~A., \& Wolfire, M.~G.\ 2009, \apj, 706, 1594 

\bibitem[Neufeld et al.(2010]{neufeld2010a}  Neufeld, D.~A.  et al., this volume

\bibitem[Neufeld et al.(2005)]{2005ApJ...628..260N} Neufeld, D.~A., 
Wolfire, M.~G., \& Schilke, P.\ 2005, \apj, 628, 260 

\bibitem[Neufeld et al.(1997)]{1997ApJ...488L.141N} Neufeld, D.~A., 
Zmuidzinas, J., Schilke, P., \& Phillips, T.~G.\ 1997, \apjl, 488, L141 

\bibitem[Neufeld et al.(1997)]{Ne}Neufeld, D.~A.\ et al.\, this volume

\bibitem[Nolt et al.(1987)]{1987JMoSp.125..274N} Nolt, I.~G., et al.\ 1987, 
Journal of Molecular Spectroscopy, 125, 274 

\bibitem[Olofsson et 
al.(2003)]{2003A&A...402L..47O} Olofsson, A.~O.~H., et al.\ 2003, \aap, 402, L47 

\bibitem[Ott 2010]{ott10} Ott, S. 2010, in ASP Conference Series, Astronomical Data Analysis 
Software and Systems XIX, Y. Mizumoto, K.-I. Morita, and M. Ohishi, eds., in press 

\bibitem[Persson et 
al.(2007)]{2007A&A...476..807P} Persson, C.~M., et al.\ 2007, \aap, 476, 807 

\bibitem[Pickett et al.(1998)]{JPL-catalog}
Pickett, H.~M., Poynter, R.~L., Cohen, E.~A., et al.
1998, J. Quant.  Spectrosc. Radiat. Transfer, 60, 883

\bibitem[Pilbratt et al. (2010)]{Pi} Pilbratt, G.\ et al.\, this volume

\bibitem[Reese et 
al.(2005)]{2005A&A...430.1139R} Reese, C., Stoecklin, T., Voronin, A., \& Rayez, J.~C.\ 2005, \aap, 430, 1139 

\bibitem[Scoville et al.(1983)]{1983ApJ...275..201S} Scoville, N., 
Kleinmann, S.~G., Hall, D.~N.~B., \& Ridgway, S.~T.\ 1983, \apj, 275, 201 

\bibitem[Snell et al.(2000)]{2000ApJ...539L..93S} Snell, R.~L., et al.\ 
2000, \apjl, 539, L93 

\bibitem[Tauber et al.(1991)]{1991ApJ...375..635T} Tauber, J.~A., 
Goldsmith, P.~F., \& Dickman, R.~L.\ 1991, \apj, 375, 635 

\bibitem[Ungerechts et al. (1997)]{1997ApJ...482..245U} Ungerechts, H., 
Bergin, E.~A., Goldsmith, P.~F., Irvine, W.~M., Schloerb, F.~P., 
\& Snell, R.~L.\ 1997, \apj, 482, 245 

\bibitem[Wright et al. (2000)]{wright2000} Wright, C.M., van Dishoeck,
  E.F., Black, J.H., et al. 2000, A\&A, 358, 689

\bibitem[Xe et al. (2008)]{xu08} Xu, L.-H., et al. 2008, J. Mol. Spectrosc. 251, 305.

\end{thebibliography}
\end{document}